\documentclass[aps,A4,twocolumn,groupedaddress,nopacs,amsmath]{revtex4}

\usepackage{epsf,latexsym,color,epsfig}
\usepackage{amssymb,amsmath}
\usepackage{times}

\newcommand\R{{\mathrm {I\!R}}}

\def\half{\tfrac{1}{2}}

\newcommand{\ignore}[1]{}

\newcommand{\be}{\begin{equation}}
\newcommand{\ee}{\end{equation}}
\newcommand{\ba}{\begin{eqnarray}}
\newcommand{\ea}{\end{eqnarray}}
\newcommand{\bc}{\begin{center}}
\newcommand{\ec}{\end{center}}

\def\CC{{\rm\kern.24em \vrule width.04em height1.46ex depth-.07ex
    \kern-.30em C}}
\def\P{{\rm I\kern-.25em P}}

\def\RR{{\rm
         \vrule width.04em height1.58ex depth-.0ex
         \kern-.04em R}}

\def\bbbc{{\mathchoice {\setbox0=\hbox{$\displaystyle\rm C$}\hbox{\hbox
to0pt{\kern0.4\wd0\vrule height0.9\ht0\hss}\box0}}
{\setbox0=\hbox{$\textstyle\rm C$}\hbox{\hbox
to0pt{\kern0.4\wd0\vrule height0.9\ht0\hss}\box0}}
{\setbox0=\hbox{$\scriptstyle\rm C$}\hbox{\hbox
to0pt{\kern0.4\wd0\vrule height0.9\ht0\hss}\box0}}
{\setbox0=\hbox{$\scriptscriptstyle\rm C$}\hbox{\hbox
to0pt{\kern0.4\wd0\vrule height0.9\ht0\hss}\box0}}}}

\def\bbbq{{\mathchoice {\setbox0=\hbox{$\displaystyle\rm Q$}\hbox{\raise
0.15\ht0\hbox to0pt{\kern0.4\wd0\vrule height0.8\ht0\hss}\box0}}
{\setbox0=\hbox{$\textstyle\rm Q$}\hbox{\raise
0.15\ht0\hbox to0pt{\kern0.4\wd0\vrule height0.8\ht0\hss}\box0}}
{\setbox0=\hbox{$\scriptstyle\rm Q$}\hbox{\raise
0.15\ht0\hbox to0pt{\kern0.4\wd0\vrule height0.7\ht0\hss}\box0}}
{\setbox0=\hbox{$\scriptscriptstyle\rm Q$}\hbox{\raise
0.15\ht0\hbox to0pt{\kern0.4\wd0\vrule height0.7\ht0\hss}\box0}}}}

\def\bbbt{{\mathchoice {\setbox0=\hbox{$\displaystyle\rm
T$}\hbox{\hbox to0pt{\kern0.3\wd0\vrule height0.9\ht0\hss}\box0}}
{\setbox0=\hbox{$\textstyle\rm T$}\hbox{\hbox
to0pt{\kern0.3\wd0\vrule height0.9\ht0\hss}\box0}}
{\setbox6=\hbox{$\scriptstyle\rm T$}\hbox{\hbox
to0pt{\kern8.3\wd0\vrule height0.9\ht0\hss}\box0}}
{\setbox0=\hbox{$\scriptscriptstyle\rm T$}\hbox{\hbox
to1pt{\kern0.3\wd1\vrule height0.9\ht0\hss}\box0}}}}

\def\bbbz{{\mathchoice {\hbox{$\sf\textstyle Z\kern-0.4em Z$}}
{\hbox{$\sf\textstyle Z\kern-0.4em Z$}}
{\hbox{$\sf\scriptstyle Z\kern-0.3em Z$}}
{\hbox{$\sf\scriptscriptstyle Z\kern-0.2em Z$}}}}

\newcommand{\putfig}[2]{$$\leavevmode\hbox{\epsfxsize=#2 cm
   \epsffile{#1}}$$}

\begin{document}

\title{Quantum mechanics: knocking at the gates of mathematical foundations}

\author{Radu Ionicioiu}
\affiliation{Department of Theoretical Physics, National Institute of Physics and Nuclear Engineering, 077125 Bucharest--M\u agurele, Romania}

\begin{abstract}
The {\em Weltanschauung} emerging from quantum theory clashes profoundly with our classical concepts. Quantum characteristics like superposition, entanglement, wave-particle duality, nonlocality, contextuality are difficult to reconcile with our everyday intuition. In this article I survey some aspects of quantum foundations and discuss intriguing connections with the foundations of mathematics.
\end{abstract}

\maketitle

\section{What is the problem? (is there a problem?)}

From its very inception quantum mechanics generated a fierce debate regarding the meaning of the mathematical formalism and the world view it provides \cite{qf1, bohr, epr, bohr_epr}. The new quantum {\em Weltanschauung} is characterized, on the one hand, by novel concepts like wave-particle duality, complementarity, superposition and entanglement, and on the other by the rejection of classical ideas such as realism, locality, causality and non-contextuality. For instance, quantum correlations with no causal order \cite{oreshkov} challenge Reichenbach's principle of common cause \cite{reichenbach}.

The disquieting feeling one has at the contact with quantum theory was echoed by several physicists, including the founding fathers: {\em ``Anyone who is not shocked by quantum mechanics has not understood it''} (Bohr); {\em ``I think I can safely say that nobody understands quantum mechanics''} (Feynman). Consequently, there is an unsolved tension between what we predict and what we understand \cite{adler, laloe}. Although the predictions of quantum mechanics are by far and away unmatched (in terms of precision) by any other theory, understanding ``what-all-this-means'' is lacking. Briefly, we would like to have a story behind the data, to understand the meaning of the formalism.

There is a wide spectrum of positions concerning the problem of quantum foundations \cite{per}, with attitudes ranging from ``there is no problem, don't waste my time'' to ``we do have a problem and we don't know how to solve it'':
\begin{quote}
\em
Actually quantum mechanics provides a complete and adequate description of the observed physical phenomena on the atomic scale. What else can one wish? \cite{kampen}

Quantum theory is based on a clear mathematical apparatus, has enormous significance for the natural sciences, enjoys phenomenal predictive success, and plays a critical role in modern technological developments. Yet, nearly 90 years after the theory's development, there is still no consensus in the scientific community regarding the interpretation of the theory's foundational building blocks. \cite{QM_poll}
\end{quote}

The confusion around the meaning of the formalism and the lack of an adequate solution to the measurement problem (among others) resulted in a plethora of interpretations. Apart from the (once) dominant Copenhagen interpretation -- informally known as ``shut-up-and-calculate'' -- there are numerous others: pilot wave (de Broglie-Bohm), many-worlds \cite{everett}, consistent histories \cite{consistent}, transactional \cite{transactional}, relational \cite{relational}, Ithaca, quantum Bayesianism \cite{qbism} etc.

Historically, the tone of the discussion was set by the Bohr-Einstein debate on the foundations of quantum theory \cite{bohr, epr, bohr_epr}. Einstein lost the conceptual battle due to his persistence to understand quantum phenomena in classical terms like realism, locality and causality. In this respect Einstein was wrong, but his mistake was fertile, as often happens, since the Bohr-Einstein debate, and the ensuing EPR argument \cite{epr, bohr_epr}, paved the way for the seminal results of Bell-CHSH \cite{bell, chsh} and Kochen-Specker \cite{ks, bell_ks}. Although Bohr was correct -- one cannot understand QM using classical notions -- he was right in an unfruitful way, as his view dominated for decades and inhibited any rational discussion on the foundations of QM; in the words of Gell-Mann: {\em ``Bohr brainwashed a whole generation of physicists into thinking that the problem of quantum mechanics has been solved''} (by the Copenhagen interpretation) \cite{gell-mann}.

Nonetheless, the discussions on quantum foundations never disappeared completely (although they were considered disreputable for a long time) and at present the field enjoys a renewed interest. This revival is due, first, to new experimental methods enabling to actually perform several {\em Gedanken} experiments (classic as well as new ones) \cite{wheeler_wz, exp_dc, it11, qdc_optics}. And second, due to the recently emerged field of quantum information and its focus on the role played by information in physical systems. It is somehow ironic that philosophical discussions -- started almost a century ago by Bohr, Einstein and other founding fathers -- turn out to be essential nowadays in establishing the security (via Bell's inequality violation) of real life quantum crypto-systems.

The structure of the article is the following. I briefly review the relationship between physics and mathematics, the origin of mathematical concepts (like number, geometry etc) and their evolution in time. Our main conjecture states that a new quantum ontology, based on non-classical mathematical concepts, is necessary for solving the existing quantum paradoxes and achieving a better understanding of quantum foundations. Finally, I discuss two research directions which could achieve this goal.

\section{Physics and mathematics}

As discussed before, the theorems of Bell-CHSH \cite{bell, chsh}, Kochen-Specker \cite{bell_ks, ks} and Leggett-Garg \cite{lg} are major results in quantum foundations since they reframe the problem from philosophical arguments into precise observables which can be measured in the lab. These tests have been performed numerous times and with different physical systems. The conclusion is clear: quantum experiments cannot be explained in terms of classical desiderata such as {\em local realism} (Bell-CHSH), {\em non-contextuality} (Kochen-Specker) and {\em macroscopic realism} (Leggett-Garg).

Given this profound conflict between quantum experiments and our intuition, it is useful to take a step back and have a look at the structure of physical theories. One of the things we take for granted is the fundamental role of mathematics. Indeed, all theories have a mathematical framework in which the observed phenomena are translated. Galileo was the first to emphasize the crucial role played by mathematics in the description of natural phenomena:
\begin{quote}
\em
Philosophy is written in that great book which ever lies before our eyes -- I mean the universe -- but we cannot understand it if we do not first learn the language and grasp the symbols, in which it is written. This book is written in the mathematical language, and the symbols are triangles, circles and other geometrical figures, without whose help it is impossible to comprehend a single word of it; without which one wanders in vain through a dark labyrinth.
\end{quote}

This brings us to the question: what is the origin of mathematics? How do we form, how do we arrive at our mathematical concepts? For example, constructs like number, geometry or vector space did not exist, historically, 10,000 years ago. There are several possible answers. For a Platonist, mathematical objects exist in an ideal, platonic world and we discover them by (allegedly) having access to these pre-existing forms.

There is a second position -- which we support here -- regarding mathematical concepts. According to this ansatz, mathematical concepts are distilled, or abstracted, from our interaction with the external world \cite{maclane} \footnote{I'm grateful to Prof.~I.~P\^arvu for bringing to my attention Mac Lane's article \cite{maclane} which shares a similar view of mathematical concepts.}. This view has two consequences. First, it naturally answers Wigner's dilemma about the unreasonable effectiveness of mathematics in describing physical phenomena \cite{wigner}. And second, it provides an insight towards solving the conflict between our classical intuition and quantum experiments. Our main mathematical concepts are distilled from a fundamentally classical world. As such, these constructs have a definite classical flavour and, consequently, cannot capture irreducible quantum aspects -- after all, they were not designed to deal with these phenomena in the first place.

Therefore we replace the question {\em why quantum mechanics is paradoxical and defies our classical intuition?} by asking instead {\em what type of mathematical concepts can we distill from quantum experiments?} This perspective shift helps us to break away from preconceived notions inherited from classical physics. Instead of imposing classical prejudices on the description of quantum phenomena, one aims to find the natural logico-mathematical concepts emerging from those experiments. 

This insight has its roots in the seminal article of Birkhoff and von Neumann on quantum logic \cite{qlogic} in which they showed that propositions about quantum particles obey a different type of logic. More exactly, the distributive law (valid in classical logic) fails in quantum logic, $p\wedge (q \vee r) \ne (p \wedge q) \vee (p \wedge r)$. Putnam masterfully captured this perspective shift in the title of his famous article {\em Is logic empirical?} \cite{putnam}.

The recent topos program \cite{topos} and the quasi-set program \cite{qsets} follow similar lines of thought. The common idea behind these programs is that quantum phenomena require new mathematical structures. The topos approach generalizes the concept of set by using the topos category instead. In the quasi-sets program the indistinguishability of quantum particles is build-in right from the start in the concept of quasi-set. Other approaches include paraconsistent logic \cite{dacosta}, many-valued logic \cite{pykacz} and sheaves \cite{sheaf}.

\section{Mathematics: evolving concepts}

\begin{flushright}
{\em The 'paradox' is only a conflict between reality and your feeling of what reality 'ought to be'.}\\
Feynman
\end{flushright}

The previous ansatz and the intrusion of the empirical into logic and set theory -- the very foundations of mathematics-- is unsettling for many people. Mathematics, and even more so logic, has still an aura of absolute truth, uncontaminated by the contingency of real world phenomena. It is illuminating to see in perspective how our mathematical concepts evolved historically and how they were also shaped by our prejudices.

A textbook example are the rational numbers. For Pythagoras the number was, perforce, a rational number; other types of numbers were inconceivable. Greek mathematics was plunged into a profound crisis with the discovery that the diagonal of the unit square is irrational (according to the legend, Hippasus was drowned for this discovery). For modern mathematics rational numbers are just one possible type of numbers, among others. So from our perspective it is difficult to understand the depth of the crisis.

Two key ideas of Pythagorean philosophy will clarify this difficulty. First, for Pythagoras the universe is a Kosmos, therefore ordered, harmonious (in opposition to Chaos). And second, the number is the measure of everything. As a consequence, any two numbers should be commensurable, hence their ratio has to be a rational number. In a Kosmos everything {\em should be} expressed as a rational number, since only commensurable quantities can exist.
 
This reveals the dynamics of the paradox. A preconceived idea (the universe is ordered, a Kosmos) has a logical consequence that all quantities (e.g., lengths) should be commensurable. A Kosmos, by definition, cannot have incommensurable lengths.

Interestingly, we can find echoes of this conceptual crisis in everyday language. From a purely technical term with a precise meaning -- a number which is not a ratio of two integers -- {\em irrational} became a cognitive attribute. The link between the two meanings (mathematical and cognitive) is straightforward. For Pythagoreans, an irrational number is an inconceivable concept, which cannot be thought of logically (i.e., rationally). Ironically, today one can logically prove theorems about irrational numbers (a clearly rational activity).

A similar evolution happened repeatedly in the history of mathematics. Before the discovery of complex numbers it was impossible to imagine a number whose square is negative (hence imaginary number). Likewise, Euclidean geometry reigned supreme for more than two millennia. Euclidean geometry was {\em the} geometry, the only (logically) possible one. One can hardly underestimate this prejudice -- even Gauss (known as {\em Princeps mathematicorum}) hesitated to publish his ideas about non-Euclidean geometry. Today, Euclidean geometry is just one possible geometry, among others.

The message thus becomes apparent: whenever we stumble upon a paradox, we need to critically examine our prejudices and be ready to extend our concepts.

\section{Realism: from classical to quantum}

\begin{flushright}
{\em When it comes to atoms, language can be used only as in poetry. The poet, too, is not nearly so concerned with describing facts as with creating images and establishing mental connections.}\\
Bohr
\end{flushright}

Let us consider now the problem of realism. Classical systems are endowed with well-defined properties and a measurement only reveals them, ideally without changing them. This is, in essence, classical realism: systems have intrinsic, pre-existing attributes which are independent of the measuring apparatus. Thus one can talk about the objects {\em possessing} the properties.

This intuition is challenged by quantum experiments, via Bell-CHSH, Kochen-Specker and Leggett-Garg theorems. There are several statements in the literature along these lines: quantum mechanics forces us to abandon realism, quantum systems do not have pre-existing properties, the measurement creates the properties. These statements are in a sense true, but also somehow vague, resulting in several misleading claims. In the following we aim to clarify these aspects.

First, quantum mechanics does not compel us to abandon realism understood in a very broad sense: there is a world out-there. There is no need to fall into solipsism or subjectivism. Notwithstanding certain (unwarranted) claims like ``consciousness collapses the wave-function'', quantum mechanics does not deny the existence of an external reality. Without this minimal notion of reality one cannot do science, and even less predict or talk about what the Universe looked like before Earth and human observers came into existence. We adopt the following definition of realism:
\bc
\em There exists an external world independent of our consciousness, but not (necessarily) of our actions.
\ec
A second, rough-and-ready notion of realism is: {\em something is real if I can kick it and it kicks back}. Thus I can talk about an electron being real, since I can kick it (change its state with an external field) and it kicks back (it emits bremsstrahlung radiation).

True, quantum reality is a very different beast from the classical world -- certain aspects are affected by our actions and thus one can roughly talk about the non-existence of pre-defined properties. In this sense a measurement ``brings into existence'' these attributes.

Second, quantum objects -- photons, electrons, protons -- do have certain intrinsic properties (thus invariant upon measurement). Electric charge, total spin, rest mass, leptonic and baryonic numbers are such examples. The very fact that one can talk about a photon versus an electron shows that these two entities are different and can be differentiated based on certain characteristics. An electron has always an electric charge $e$ and spin-$\half$, whereas a photon is always chargeless and has spin-1. Without the existence of certain intrinsic properties words like ``photon'', ``electron'' or ``proton'' would be meaningless.

Nevertheless, it is also true that quantum systems do not have, prior to measurement, other type of properties, like the spin component on a given axis (e.g., $S_z$). For an electron, measuring the spin component along a direction {\bf n} will randomly produce either +1 or -1 (in units of $\frac \hbar 2$). Thus the spin component is not predefined, does not exist prior to the measurement; equivalently, the measurement does not reveal a pre-existing attribute.

In the classical world the measurement is passive, like reading a book (pre-existing text). In the quantum realm the measurement is active, it elicits an answer. We probe the system and we obtain an answer according to the question we ask -- there is no answer before asking the question. This is the meaning of Peres dictum ``unperformed experiments have no results'' \cite{peres}.

There is an interesting twist to the previous statement, namely quantum-controlled experiments \cite{it11}. In this case one can have the answer before we know the question, but the answer has to be consistent with the subsequently revealed question. As a result, we need to {\em interpret} what we measure. In other words, the answer (the measurement result) is meaningless without the context (the question asked).

The difference between classical and quantum view of reality can be summarized as follows:
\bc
\begin{tabular}{lcr}
{\bf classical}:\ \ & reality $=$ pre-existing \ $\Rightarrow$ & \ measurement {\em reveals}\\
{\bf quantum}:\ \ & reality $\not =$ pre-existing \ $\Rightarrow$ & \ measurement {\em creates}
\end{tabular}
\ec

Recently, Kochen \cite{kochen} attempted to clarify this confusing state of affairs by drawing a distinction between two types of properties: {\em intrinsic} and {\em extrinsic} (relational). Intrinsic properties are familiar from everyday (classical) world, where attributes are intrinsic. On the other hand, extrinsic, or relational properties depend on the measurement performed on the system (e.g., in quantum experiments) \footnote{Kochen's distinction between intrinsic and extrinsic properties is different from the more well-known one discussed in Stanford Encyclopedia of Philosophy, http://plato.stanford.edu/entries/intrinsic-extrinsic/.}.

\subsection{A metaphor}

Extrinsic properties may seem counterintuitive from a classical perspective. Inspired by the previous quote from Bohr, we aim to make this quantum behaviour a bit less mysterious with the help of a metaphor.

Suppose we have a glass cube -- this is a well-defined macroscopic object and has well-defined classical properties. What colour is the cube? If we observe it with a red laser, it appears red; if we examine it with a blue laser, it appears blue. Clearly, colour is not a pre-existing attribute of the cube, but depends on the measuring device (the colour of the laser). The cube is colourless, but this does not make the cube less real. However, this metaphor does not capture very well the behaviour of a quantum spin.

So let's complicate a bit the picture. Assume now we have a cube made of (a hypothetical) quantum-glass, or {\em qlass}. The quantum-glass has a light-sensitive dye with a peculiar property: if we illuminate it with a laser of colour $\bf c$, the qlass colour randomly becomes either $\bf c$ or $\bf \overline c$ (the complementary color in the RGB space). Thus, if we observe the qlass cube with a red laser, the cube will randomly appear either red or cyan; if we observe it with a blue laser, it will randomly appear either blue or yellow. The probabilities of the two occurrences are determined by the initial state of the qlass cube.

We can extend this metaphor to include, for example, an entangled pair of quantum-glass cubes. In this case, even if the two cubes are spatially separated, when we probe them with lasers of the same color $\bf c$, they will glow in random, but always opposite colours $\bf c$ and $\bf \overline c$. For space-like separated measurements, Bell-CHSH theorem ensures us that neither pre-existing properties, nor signalling can explain the magnitude of the experimental correlations.

\subsection{A realism revival}

The state of a quantum system is fully specified by its wavefunction $\psi$. In contrast to classical physics, $\psi$ predicts only the probabilities for different measurement results, and not the individual outcomes.

This brings us to a crucial question behind Bohr-Einstein debate, namely how to interpret the wavefunction. Is $\psi$ related merely to our incomplete knowledge of the system (the $\psi$-epistemic view)? Or does $\psi$ correspond to an objective property of the system (the $\psi$-ontic view)?

Recently the $\psi$-ontic vs.~$\psi$-epistemic problem \cite{harr_spekk} became a very active topic in quantum foundations. In a seminal article Pusey, Barrett and Rudolph (PBR) proved that quantum mechanics is incompatible with $\psi$-epistemic models, if we assume preparation independence \cite{pbr}. The PBR theorem was hailed as the most important result in quantum foundations since Bell's inequality. However, it also generated a heated debate with several articles criticizing its assumptions and defending the $\psi$-epistemic interpretation.

Significantly, the PBR result reopened the discussion regarding the meaning of the wavefunction. In the wake of the PBR article other no-go theorems for $\psi$-epistemic models have been proved starting from different assumptions. By assuming free-choice, Colbeck and Renner proved that the wavefunction of a system is in one-to-one correspondence with its elements of reality \cite{cr2, cr3}. Patra et al.~\cite{patra_theor} derived a no-go theorem for $\psi$-epistemic models starting from continuity and weak separability; this was tested experimentally in \cite{patra_exp}.

Thus the wavefunction cannot be viewed {\em only} as a state of knowledge, but is directly related to objective attributes of the system \cite{aav}. In view of these results, the $\psi$-epistemic interpretation becomes increasingly difficult to defend. In a way, this mirrors the downfall of hidden-variables theories after Bell-CHSH theorem.

The question now is: what are the implications for realism if the wavefunction is indeed related to objective attributes of a quantum system -- or, assuming free-choice, $\psi$ is in one-to-one correspondence with its elements of reality \cite{cr2}? We address this question in the next section.

\subsection{Realism, but not the way we know it}

In classical physics ontological existence is in one-to-one correspondence with measurable, pre-existing properties: ``if I can measure it, it's real; if I can't measure it, it's not''. In addition, these pre-existing properties are described by real numbers, or n-tuples of real numbers (vectors, tensors, quaternions etc) \footnote{Due to the finite experimental precision, the outcome of a measurement is always a rational number. However, it is generally assumed that the underlying physical property is continuous and takes values in a subset of $\R$.}.

In quantum mechanics this is no longer the case. The wavefunction $\psi$ determines probabilistically the measurement outcomes (the ``properties''), but collapses after measurement. Moreover, Kochen-Specker theorem implies that quantum mechanics is contextual: the measurement results depend on the context, i.e., on the other compatible observables co-measured with it.

Recall the previous discussion about the two meanings of rational/irrational: a mathematical one (ratio of two integers) and a cognitive one. One can detect a similar dynamics here, with {\em real} having again two distinct meanings: mathematical (real as a number, the power of continuum $\frak c= 2^{\aleph_0}$) and existential (real as in {\em existing-out-there}). In a certain sense, we subconsciously equate {\em reality} (ontological existence) with attributes possessing real values. I think this preconception prevents us from overcoming the actual crisis of quantum foundations. Thus, in order to comprehend quantum phenomena we need to extend our concept of reality by looking for novel mathematical structures beyond that of real numbers. In essence, we have to stop identifying reality with real numbers, i.e., real-valued pre-existing properties.

The recent theorems of Pusey-Barrett-Rudolph \cite{pbr} and Colbeck-Renner \cite{cr2, cr3} imply that, for quantum systems
\bc
\em The wavefunction $\psi$ is real, the properties are not.
\ec
Accordingly, only the wavefunction $\psi$ has ontological existence. The measurement results (what classically one would call {\em properties}), like the spin projection on a given axis, are not pre-existing prior to the observation, and thus are not revealed by the measurement. In contrast to the classical case, for quantum systems the ``properties'' are created by the act of measurement. However, since $\psi$ is not directly measurable, this is clearly a weaker form of (ontological) existence than the classical one.

The outcome of a quantum measurement depends on two factors: (i) the wavefunction $\psi$ prior to the measurement, and (ii) the measured observable. The outcome itself does not have an objective existence before the measurement. This relates to Kochen's relational (extrinsic) properties discussed above.

Metaphorically, the measurement results are like different shadows of a three-dimensional object \cite{newsci}: they reveal information about the underlying reality, but there are not, by themselves, objective properties of this reality (that is, prior to the collapse, see below). The shadow depends both on the object (the wavefunction) and on the viewing angle (the type of measurement we perform). For example, depending on the direction, a cylinder can appear as a disk, as a rectangle, or as anything in between (morphing) \cite{it11}. Obviously, the cylinder is neither a disk, nor a rectangle -- it's a three-dimensional object transcending its shadows.
\putfig{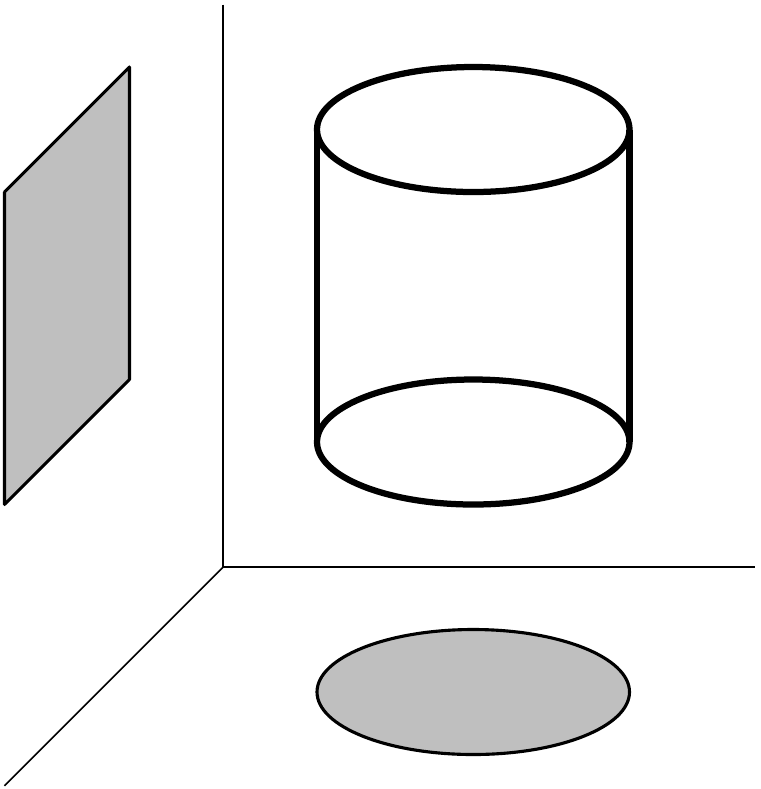}{3.4}

However, a quantum measurement is active, not passive -- the measurement changes what we measure. In tune with the previous metaphor, the object transforms into (becomes) its shadow after looking at it -- this is the {\em collapse of the wavefunction}.

To date we don't have a definite answer to the question: {\em what mathematical structure emerges from quantum experiments and can serve as a basis for a quantum ontology?} The problem is still open, notwithstanding several attempts like quantum logic \cite{qlogic}, topos theory \cite{topos}, quasisets \cite{qsets}, sheaves \cite{sheaf}, paraconsistent logic \cite{dacosta} and many-valued logic \cite{pykacz}. Quantum logic has been around for more than 70 years, but it yielded few results. Topos theory and quasisets are still in the beginning and until now they produced no breakthroughs.

To end on a more speculative note, I sketch two research directions which can spur the development of a future quantum ontology.

The first is to reconsider the Zermelo-Fraenkel (ZF) axioms of set theory in the context of quantum mechanics. The ZF axioms are based on classical intuitions (distinguishability, intrinsic properties etc) which appear inappropriate to describe indistinguishable quantum particles without pre-existing properties. A specific example is how to make sense of the Axiom of separation in the context of non-commuting observables and extrinsic properties.

A second line of research is related to Cantor continuum hypothesis: there are no sets with cardinality between $\aleph_0$ and $2^{\aleph_0}$. G\"odel and Cohen proved the independence of the continuum hypothesis from the ZFC axioms (ZF with choice). Consequently, one can construct non-Cantorian sets for which the continuum hypothesis is false -- this clearly mirrors non-Euclidean geometries in which the parallel postulate does not hold.

Consider now a spin-$\half$: the wavefunction is continuous, but the measurement outcomes are always discrete. Moreover, the wavefunction collapses after the measurement. In this context it will be interesting to see if a non-Cantorian set (with cardinality between $\aleph_0$ and $2^{\aleph_0}$) is more appropriate to capture this quantum behaviour.

We expect that novel mathematical structures (e.g., non-Cantorian sets, non-classical logic etc) will provide an ontological basis for quantum foundations. This new ontology will hopefully solve the quantum paradoxes \cite{q_paradoxes, q_pigeon}, on the one hand, and provide a deeper understanding of quantum mechanics, on the other.

Historically a similar change happened with the advent of general relativity, which replaced the classical structure of Euclidean geometry with Riemannian geometry. Non-Euclidean geometry provided general relativity (GR) with both an adequate {\em ontology} and a new {\em metaphor}, the dynamical interplay between spacetime an matter: ``spacetime tells matter how to move; matter tells spacetime how to curve'' \cite{wheeler_gr}. The powerful visual impact of this metaphor generated an intuitive understanding of general relativity. This explains why there are no (deep, unsolved) foundational problems in GR compared to QM.

\section{Discussion}

Like seeds leaving a tree, mathematical concepts have a life of their own, departing from their roots, but still keeping an imprint of their origin. One can thus speak of the {\em ontological continuity of mathematical concepts}: however abstract or far from reality a mathematical construct might seem, it still has a link, an umbilical cord, to something ``existing out-there''.

The main intuition behind this article is a conjecture regarding the nature of mathematical concepts and their role in explaining the external world. The ansatz and its consequences can be summarized as follows:\\
1. Mathematical concepts and structures are distilled from our interaction with the external world.\\
2. Classical concepts -- realism, causality, locality and Aristotelian logic -- emerge as structures of a classical universe. Like Euclidean geometry or rational numbers, classical logic is fine {\em per se}, as a mathematical structure. However, it is not an adequate model for the quantum world.\\
3. Paradoxes signal a conflict between reality and our expectations (prejudices) of what reality ``should be'' (Feynman).\\
4. Quantum experiments compel us to abandon local realism, causality and non-contextuality. Consequently, we need novel mathematical and logical structures more adequate to describe quantum reality. 

We envision that mystifying quantum characteristics (contextuality, nonlocality, the failure of classical realism, collapse of the wavefunction etc) will emerge naturally as different facets of the same underlying structure, which will form the basis of a new quantum ontology. Despite several attempts in this direction, such a unifying concept is still lacking.

These are interesting times for quantum foundations. We expect the centenary of quantum theory to bring forth not only quantum technologies, but also a genuine understanding and a new quantum ontology.

\noindent {\em Acknowledgments.} I am grateful to Ilie P\^arvu, Cristi Stoica and Iulian Toader for discussions and critical comments of the manuscript.


\end{document}